\documentclass[12pt]{article}
\usepackage{epsfig,amsfonts,amssymb}
\usepackage{hyperref}
\pdfoutput=1

\usepackage{cite}
\usepackage{cite}

\usepackage{comment}

\def\ZZZ{{\hbox{ Z\kern-1.6mm Z}}}
\def\RRR{{\hbox{ R\kern-2.4mm R}}}
\def\CCC{{\hbox{ C\kern-2.0mm C}}}
\def\zzz{{\hbox{z\kern-1mm z}}}

\newcommand{\qeq}{{\hbox{=\kern-2.3mm ? \kern.5mm }}}
\renewcommand{\qeq}{=}

\newcommand{\eps}{\epsilon}

\newcommand{\OO}{{\cal O}}

\newcommand{\wt}{\widetilde}
\newcommand{\wh}{\widehat}

\newcommand{\be}{\begin{equation}}
\newcommand{\ee}{\end{equation}}
\newcommand{\ben}{\begin{eqnarray}\displaystyle}
\newcommand{\een}{\end{eqnarray}}

\newcommand{\refb}[1]{(\ref{#1})}
\newcommand{\p}{\partial}
\newcommand{\sectiono}[1]{\section{#1}\setcounter{equation}{0}}

\newcommand{\gsim}{\stackrel{>}{\sim}}
\newcommand{\lsim}{\stackrel{<}{\sim}}

\def\one{{\hbox{ 1\kern-.8mm l}}}
\def\zero{{\hbox{ 0\kern-1.5mm 0}}}

\newcommand{\bea}[1]{\begin{eqnarray}\label{#1} }
\newcommand{\eea}{\end{eqnarray}}

\newcommand{\eqref}{\refb}

\usepackage{bm}
\usepackage[table]{xcolor}

\begin{document}

\baselineskip 24pt

\begin{center}

{\Large \bf How to Create a Flat Ten or Eleven 
Dimensional Space-time in the Interior
of an Asymptotically Flat Four Dimensional String Theory}

\end{center}

\vskip .6cm
\medskip

\vspace*{4.0ex}

\baselineskip=18pt

\centerline{\large \rm Ashoke Sen}

\vspace*{4.0ex}

\centerline{\large \it International Centre for Theoretical Sciences - TIFR 
}
\centerline{\large \it  Bengaluru - 560089, India}


\vspace*{1.0ex}
\centerline{\small E-mail:  ashoke.sen@icts.res.in}

\vspace*{5.0ex}

\centerline{\bf Abstract} \bigskip

By taking large mass and charge limit
of a black hole in string theory we can create arbitrarily
large regions where the space-time  
is approximately flat, but the moduli fields take values different from their 
asymptotic values. In this paper we describe a special case of this where 
black hole solutions in a four dimensional string theory, in the large
mass and charge limit, can have 
an arbitrarily large region outside the horizon 
where  a local observer  will experience type IIA string theory in flat
ten dimensional space-time. The curvature and other field strengths
remain small everywhere between the asymptotic four dimensional
observer and the ten dimensional region. By  
going to a different region of space, we can also get a large region where
a local observer experiences M-theory in flat eleven dimensional space-time.
By taking another solution in the same theory, one can create 
an arbitrarily large region
where  a local observer  will experience type IIB string theory in flat
ten dimensional space-time.

\vfill \eject

\tableofcontents

\sectiono{Introduction} \label{sintro}

In a previous paper\cite{2502.07883} we had described a general procedure by which
we can start with a string theory with a given set of values of the moduli fields at
infinity and produce
a background in which we have a large locally flat region where the moduli take 
different values. Thus an asymptotic observer can send out experimentalists to these
regions to learn about the spectrum and S-matrix in a `different string theory'. This in fact
demonstrates that the apparently different string theories
characterized by different values of the
moduli\cite{2501.17697} are not different theories but
different vacua of the same theory.

One example of this was discussed in \cite{2502.07883} where it was shown how a
large black hole in ten dimensional type IIA string theory carrying D0-brane charge
allows us to create arbitrarily large, almost flat
regions where the string coupling remains
almost constant at any value between its asymptotic value and infinity.
A large string coupling would correspond
to M-theory in an almost flat eleven dimensional  space-time\cite{9501068,9503124}.
In this paper we describe another example where we begin with
type IIA
string theory compactified on $T^6$ and produce a background that contains a large,
locally flat region where the size of the internal $T^6$ can be made arbitrarily
large, keeping the string
coupling fixed.  
Therefore a local observer in this region will experience type IIA
string theory in ten dimensional flat space-time. 
There is another region in the same background where a local observer experiences
M-theory in eleven dimensional flat space-time.
The only price
we pay is that larger the size of this ten or eleven
dimensional region, the larger is the mass of the black hole
needed to create such a configuration. Finally, by taking a different background
in the same theory, one can construct an arbitrarily large region where a local
observer experiences type IIB string theory in ten dimensional flat space-time.

The rest of the paper is organized as follows. In section \ref{sone} we describe a background
where we have a locally flat region with large internal $T^6$, but
the string coupling in
that region is also large. This is not quite the configuration we want since we want the string
coupling to  remain small or finite. 
We remedy this problem in section \ref{stwo} by placing 
the black hole solution of section \ref{sone} in the background of a
second black hole. There we show that it is not only possible to identify a
large region where the local physics is described by type IIA string theory in ten
dimensional flat space-time, but it is also possible to identify another large region
where the local physics is described by M-theory in eleven dimensional flat
space-time. In section \ref{siib} we show, how by taking a different background
in the same theory, we can produce a large region where a local observer
experiences type IIB string theory in ten  dimensional flat
space-time with moderate or small string coupling. In section \ref{ssingle} we
give an alternative construction by which we can achieve each of the three
configurations -- type IIA in ten dimensiona, M-theory in eleven dimensions and
type IIB in ten dimensions by single black hole solutions. 
We end in section \ref{sthree} with
some comments.

\sectiono{D0-brane black hole in type IIA on $T^6$} \label{sone}

We shall consider a four dimensional string theory obtained by compactifying type IIA
string theory on $T^6$ and consider a non-rotating black hole in this theory carrying
D0-brane charge. However in order to use the results of \cite{horowitz} directly, we
shall first consider a black hole carrying D6-brane charge and then obtain the solution
carrying D0-brane charge by T-dualizing all the compact directions.
The relevant part of the ten dimensional action is given by:
\be\label{e1.0}
\int d^{10}x \sqrt{-\det G} \left[ e^{-2\phi_{10}}
\left\{R_G + 4 G^{\mu\nu} \p_\mu\phi_{10} \p_\nu\phi_{10} 
\right\} - G^{\mu\rho} G^{\nu\sigma} F_{\mu\nu} F_{\rho\sigma}
\right]\, ,
\ee
where $F$ is the field strength associated with the RR one form gauge field, 
$G_{\mu\nu}$ is the string metric and $\phi_{10}$ is the dilaton field. The  
D6-brane is magnetically charged under the RR one form gauge field, and 
the black hole solution carrying such charges is given by\cite{horowitz}:
\ben\label{e1.1}
F &=&  Q\, e^{-\phi_0}\, \eps_2 \nonumber\\
ds^2 &=& - \left[ 1 - \left({r_+\over r}\right)\right]\, 
\left[ 1 - \left({r_-\over r}\right)\right]^{\gamma_x-1} dt^2
+ \left[ 1 - \left({r_+\over r}\right)\right]^{-1}\, 
\left[ 1 - \left({r_-\over r}\right)\right]^{\gamma_r} dr^2\nonumber \\
&+& r^2\, 
\left[ 1 - \left({r_-\over r}\right)\right]^{\gamma_r+1} d\Omega_2^2
+ \left[ 1 - \left({r_-\over r}\right)\right]^{\gamma_x}  \, \sum_{i=4}^9 dx^i dx^i\, ,
\nonumber \\
e^{-2\phi_{10}} &=&  e^{-2\phi_0}\, 
\left[ 1 - \left({r_-\over r}\right)\right]^{\gamma_\phi}\, , \qquad r_-\le r_+\, ,
\een
where $\eps_2$ is the volume form on a unit two sphere whose metric is 
denoted by $d\Omega_2^2$ and\footnote{The case we are considering
corresponds to the $\alpha=0$, $D=4$
case of \cite{horowitz}.}
\be
\gamma_x = {1\over 2}\, , \qquad \gamma_r = {1\over 2}\, , \qquad \gamma_\phi
= -{3\over 2}\, .
\ee
$x^4,\cdots, x^9$ are coordinates along the compact directions, which we shall take to
be periodic with period $2\pi$. The $e^{-\phi_0}$ factors were not present in the solution
given in \cite{horowitz}, but we have included it here using the scaling `symmetry' 
$\phi_{10}\to \phi_{10}+\phi_0$, $F\to e^{-\phi_0}F$ under which the action scales by
$e^{-2\phi_0}$ and the equations of motion remain unchanged. The parameters
$Q$ and the mass $M$ of the black hole solution are related to $r_\pm$ via the
relations\cite{horowitz}:
\be
Q={1\over 2} (r_+r_-)^{1/2}, \qquad M = r_+ - {1\over 2} r_-\, .
\ee
The horizon of this black hole is at $r=r_+$.
Extremal limit would correspond to $r_-\to r_+$, but for now we keep $r_-$ and
$r_+$ as independent parameters.

Now any two derivative action in $d$ dimensions 
containing metric $G_{\mu\nu}$, and $p$-forms $A^{(p)}$
for different values of $p$, scales by a factor of $\lambda^{d-2}$ under the scaling
\be\label{escaling}
G_{\mu\nu}\to\lambda^2\, G_{\mu\nu}, \qquad A^{(p)} \to \lambda^p\, A^{(p)}\, .
\ee
This takes a solution to a new solution. In the new solution any invariant constructed
from these fields with two derivatives, e.g. the curvature scalar or 
$G^{\mu\rho} G^{\nu\sigma} F_{\mu\nu} F_{\rho\sigma}$ will scale as $\lambda^{-2}$.
Invariants containing larger number of derivatives will scale as larger powers of
$\lambda^{-1}$.  Thus in the large $\lambda$ limit all these invariants will approach
zero and the space-time will appear to be locally flat. 
We shall carry out this rescaling for the solution \refb{e1.1} by regarding 
this as a solution in four dimensional
theory. This means that we do not scale the internal components of the metric, rather
we regard the sizes of the internal circles as scalar moduli fields in the four dimensional
theory and hold them fixed as we scale the other components of the metric
by $\lambda^2$. 
In the solution \refb{e1.1} this will correspond to scaling $r$, $t$, $r_+$ and $r_-$
by $\lambda$.\footnote{While the scaling of $r$ and $t$ are just coordinate 
transformations, the scaling of $r_\pm$ generate genuinely different solution.}
We shall not display $\lambda$ explicitly but keep in mind that 
we work in the region where $r_\pm$ are large and $r$ is of order $r_+$. 
It is easy to check that in this case the metric is locally approximately flat and the 
magnitude of the field
strength is approximately zero. Total integrated flux however scales as $\lambda$ 
and is large.

We now define
\be
a\equiv r_+-r_-, \qquad b = r - r_+\, ,
\ee
and take $a/r_+$ and $b/r_+$ to be small but fixed as we scale $r_\pm$ to large
values. This means that we stay in the near horizon region of a large, near extremal
black hole. The local geometry is still nearly flat  as long as we take $r_+$ to be
sufficiently large. 

We now compute the scalar $\phi_{10}$ and the radii $R$ of the compact directions in this
region. They are given by
\be\label{e1.7}
e^{-2\phi_{10}} = e^{-2\phi_0} \left( {r_++b\over b+a}\right)^{3/2}\, , \qquad R=
\left( {b+a\over r_++b}\right)^{1/4}\, .
\ee
In particular the four dimensional dilaton $\phi_4$, is given by
\be\label{Pe1.7}
e^{-2\phi_4} \equiv e^{-2\phi_{10}} \, R^6 = e^{-2\phi_0}\, .
\ee
Thus $\phi_4$ does not flow and its value at $r=r_++b$ is the same as at infinity.

From \refb{e1.7} we see that for $b,a<<r_+$, the sizes of the compact directions,
measured in the string metric, is small and $e^{-2\phi_{10}}$ is large, which means that the
string coupling is small. We shall now go to the T-dual description of this geometry,
by making an $R\to \wt R= 1/R$ transformation on all the circles. During this transformation
the four dimensional dilaton $\phi_4$
remains the same, $\phi_{10}$ transforms to $\wt\phi_{10}$ accordingly, 
and the integrated flux of gauge field is reinterpreted
as the D0-brane charge of the black hole. 
The near horizon moduli now take the form:\footnote{Note that
we could get this result directly by solving the equations of motion of low energy
supergravity. The only reason for going via D6-branes is to lift the solution from
\cite{horowitz}.}
\be\label{e1.9}
\wt R = R^{-1}= 
\left( { r_++b\over b+a}\right)^{1/4}, \qquad e^{2\wt \phi_{10}} = e^{2\phi_4} 
\, \wt R^{6}
= e^{2\phi_0} \, \left( { r_++b\over b+a}\right)^{3/2}\, .
\ee
We also note that in the new duality frame, the asymptotic radii of the circles remain  
1 and the dilaton $\wt \phi_{10}$ takes value $\phi_0$.

We see from \refb{e1.9} that for 
$b,a<<r_+$, the radii $\wt R$ of the compact directions, measured in the string
metric, is large. However the string coupling $e^{\wt\phi_{10}}$ in this
region is large. Hence we still have
not reached our goal, which is to produce a region of space-time where the internal radii
can be made (arbitrarily) large and the string coupling can be kept small or finite.

To address this issue, we note that if we are 
allowed to adjust $\phi_0$, then we could
take $e^{2\phi_0}$ to be small to compensate for the large contribution from the
last factor in the expression for $e^{2\wt\phi_{10}}$ in \refb{e1.9}. 
However we would like to get a
large ten dimensional space-time in a given four dimensional string theory for which
all the asymptotic 
moduli are fixed, including the coupling $e^{\phi_0}$. 
So $\phi_0$ cannot be
adjusted to achieve our goal.
In the next section we shall describe
a different black hole solution for which we can identify arbitrarily large locally flat
regions of space-time where
the string coupling can be made arbitrarily small, keeping all other moduli finite.
Thus by placing the black hole considered in this section (the first black hole)
in the background of  the
second black hole that we shall describe in the next section, 
we can vary the effective $\phi_0$ that enters the computation
of this section by varying the position of the first black hole.
This way we'll be able to achieve our goal of creating a region where the space-time is
approximately ten dimensional and have finite ten dimensional string coupling.

\sectiono{Black hole with  winding and momentum
charge} \label{stwo}

In this section we shall analyze a non-rotating 
black hole solution in type IIA string theory on $T^6$, carrying fundamental string
winding and momentum charge along one of the circles. For simplicity we shall take the
momentum and winding charges to be equal. 
We also take the asymptotic radii of the circles to be 1 and the 
asymptotic string coupling
to be  1.
The solution is obtained from the
solutions given in \cite{9411187} and reviewed in appendix \ref{selectric} by setting
$\alpha=0$ and taking the unit vector $\vec p$ 
to be along the circle carrying the
winding and momentum charges.
The four dimensional string
metric $ds_4^2$,  
the ten dimensional dilaton $\phi_{10}$, the four dimensional dilaton
$\phi_4$, the $12\times 12$ symmetric, $SO(6,6)$
matrix valued scalar $M$ and
the non-vanishing component of the gauge fields are given by:
\ben \label{e2.1}
ds_4^2 &=& -\Delta^{-1} \rho^2 (\rho^2-2m\rho) dt^2 + \rho^2 (\rho^2-2m\rho)^{-1} d\rho^2
+ \rho^2 d\Omega_2^2 \, , \nonumber \\
e^{-2\phi_{10}} &=& e^{-2\phi_4} = \Delta^{1/2} \, \rho^{-2} \, , \nonumber \\
M &=& I_{12}\, , \nonumber \\
A_t &=& -{1\over \sqrt 2} \, m\, \rho\, \sinh\beta \, \Delta^{-1/2}\, ,
\nonumber \\
\Delta &\equiv& (\rho^2 + m\rho(\cosh\beta-1))^2\, .
\een
The horizon is at $\rho=2m$. The $M=I_{12}$
result, $I_{12}$ being the $12\times 12$ identity matrix,
implies that the radii of the compact directions remain constant at 1 for all $\rho$.
$A_t$ is the gauge field that couples to the momentum plus winding charge.
The extremal limit corresponds to $m\to 0$, $\beta\to\infty$ with $m\sinh\beta$ fixed.
We shall not take this limit for now.

The scaling that makes the metric almost locally flat and the invariant tensors almost
zero is $m\to \lambda' m$, $\rho\to \lambda'\rho$, $t\to \lambda' t$ for large $\lambda'$.
Again we shall not display the factors of $\lambda'$ explicitly but keep in mind that we work
with large $m$ and $\rho$ with $\rho/m>2$ fixed. Also we shall take $\beta$
large but fixed. In this limit the geometry becomes nearly flat and all the invariant tensors
constructed from two or more derivatives of fields become nearly zero. The dilaton $\phi_{10}$
has the form:
\be\label{e2.3}
e^{-2\phi_{10}} \simeq  m\, \rho^{-1} \cosh\beta\, .
\ee
We now see that for large $\beta$ with fixed $\rho/m >2$, 
$e^{-2\phi_{10}}$ is large and hence $\phi_{10}$ becomes
large and negative. Therefore in this region the string coupling $e^{\phi_{10}}$ 
is small. If we take the parameter
$\lambda'$ to be much larger than the parameter $\lambda$ used in the scaling in section
\ref{sone}, then the size of the black hole described in section \ref{sone}, even though large,
is small compared to the distance scale over which the fields change appreciably  in the
black hole background given in \refb{e2.1}. Thus the black hole of section \ref{sone} will
appear to be a point object from the perspective of the background \ref{e2.1} and it makes
sense to place it at any point in the background \refb{e2.1}. In particular we can place
it at $\rho\sim m$. This configuration will evolve with time, but the time scale
over which the fields will evolve will be of order $\lambda'$ that can be made
arbitrarily large.  In that case $\phi_0$ appearing in \refb{e1.9} will be given by
the $\phi_{10}$ in \refb{e2.3}.
Substituting this into \refb{e1.9} gives
\be\label{e1.9a}
\wt R = \left( { r_++b\over b+a}\right)^{1/4}, \qquad e^{2\wt \phi_{10}} 
=  {\rho\over m \cosh\beta}\, 
\, \left( { r_++b\over b+a}\right)^{3/2}\, .
\ee
Taking $a,b<<r_+$, $\rho\sim m$ and $\beta\sim \cosh^{-1} [\{r_+/(b+a)\}^{3/2}]
 >>1$, we can achieve our goal  of
getting large $\wt R$ and finite $\wt\phi_{10}$ i.e. a locally flat geometry where space-time is
ten dimensional and the string coupling is finite.

It is also interesting to explore the parameters that arise if we regard this as an
eleven dimensional 
$M$-theory compactification using the duality between type IIA string theory in
ten dimensions and circle compactification of $M$-theory\cite{9501068,9503124}. 
Let $R_M$
denote the radii of the $x^i$ directions for 
$4\le i\le 9$  in the eleven dimensional metric 
and  $R_{11}$ denote the radius of the emergent circle as we go from type IIA to
M-theory description. In this case, we have\cite{9503124}
\be
R_{11} = e^{2\wt\phi_{10}/3} = \left({\rho\over m \cosh\beta}\right)^{1/3}
 \left( { r_++b\over b+a}\right)^{1/2}\, , 
\qquad R_M = e^{-\wt\phi_{10}/3} \, \wt R = \left( {m\cosh\beta\over \rho}\right)^{1/6}
\, .
\ee
From this we see that by taking $\beta$ and  $r_+/(a+b)$  large, and
appropriately choosing their ratios, it is possible to make both $R_{11}$ and $R_M$
large. In that case we would create a large region where a local observer would
experience eleven dimensional M-theory in almost flat space-time. In fact, for the
same geometry, characterized by large values of $r_+/a$ and $\beta$, we can
vary $b\equiv r-r_+$, 
labelling the region, to have either $\wt R$ large with $\wt \phi_{10}$ small or
finite (IIA in flat space-time), or $R_{11}$ and $R_M$ large (M-theory in flat
space-time).

Since the constructions described above involve taking various combinations
large, we
shall now summarize the order of limits needed for getting the ten and eleven
dimensional regions:
\begin{enumerate}
\item Type IIA in flat ten dimensional space-time:
\be \label{erangeiia}
m >> r_+ >>> r_+/(b+a) \sim (\cosh\beta)^{2/3} >> 1,  \qquad \rho-2m \sim m\, .
\ee
We remind the reader that $a/r_+\equiv (r_+-r_-)/r_+$ is the non-extremality parameter
of the first black hole, $1/\cosh\beta$ is the non-extremality parameter
of the second black hole, $r=r_++b$ labels the region around the first
black hole where we work and $\rho$ labels the region around the second black hole
where we work. Also we have set $\hbar=1$, $c=1$, $\alpha'=1$. The $m>>r_+$ 
condition ensures that the second black hole is much larger than the first
black hole. The $m, r_+>>>
 r_+/(a+b), (\cosh\beta)^{2/3}$ condition ensures that the 
curvature and other invariants from the perspective of four dimensional 
space-time are made small enough so that even if the natural
invariants from the higher dimensional perspective carry some
powers of the sizes of the extra dimensions, these invariants still remain small.
So these inequalities should really be interpreted as taking $m$ and $r_+$ to be
larger than any power of $r_+/(a+b)$ and $(\cosh\beta)^{2/3}$ 
that might arise in the expressions for ten dimensional invariants constructed from
derivatives of the metric and other fields. This is what is meant by the $>>>$
symbol.
\item M-theory in flat eleven dimensional space-time:
\be
m >> r_+ >>> r_+/(b+a) >> (\cosh\beta)^{2/3} >> 1, \qquad  \rho-2m\sim m\, .
\ee
As in the ten dimensional case, we need to take $m$ and $r_+$ to be
larger than any power of $r_+/(a+b)$ and $(\cosh\beta)^{2/3}$ that might arise
in the expression for the eleven dimensional invariants constructed from derivatives
of metric and other fields.

\end{enumerate}

\sectiono{Black hole with purely winding charge} \label{siib}

In this section we shall describe how to construct a configuration that
contains an arbitrarily large region where the local observer experiences type IIB string
theory in flat ten dimensional space-time. 
For this we need to make use of a third black hole solution in 
type IIA string theory on $T^6$ that carries only winding charge along one of the
circles. We shall take this to be the ninth circle. This solution
can again be read out from the
general solution given in 
appendix \ref{selectric} by 
setting $\alpha=\beta$ and taking 
the vectors $\vec n$ and $\vec p$ to be directed along the same
compact direction $x^9$. 
Using \refb{e2.1app}-\refb{emr},
the four dimensional dilaton $\phi_4$ and the radius $R_9$ of the
ninth direction  takes the form:
\be \label{e2.1nn}
e^{2\phi_4} = e^{2\wh\phi_0} \, 
(1+ 2 \wh m\wh\rho^{-1} \sinh^2\alpha)^{-1/2}\, ,
\ee
\be\label{e2.1nnn}
R_9 = (1 + 2\wh m \wh \rho^{-1}  \sinh^2\alpha)^{-1/2}\, ,
\ee
where we have introduced new parameter $\wh m$ and new radial variable $\wh\rho$
to distinguish them from the variables used in section \ref{stwo}.
The other radii are kept fixed at 1.
${\wh\phi_0}$ denotes the asymptotic value of $\phi_4$ for this solution.
This is also the asymptotic value of the ten dimensional dilaton
$\phi_{10}$ for this solution since $R_9$ approaches 1
in this limit and the other radii are also fixed at 1. On the other hand at a generic
position labelled by $\wh\rho$, the ten dimensional dilaton $\phi_{10}$ takes the form
\be\label{efro1}
e^{2\phi_{10}} = e^{2\phi_4} \, R_9 =  e^{2\wh\phi_0} \,
(1 + 2\wh m \wh \rho^{-1}  \sinh^2\alpha)^{-1}\, .
\ee
The horizon of this solution is at $\wh\rho=2\wh m$. Hence we shall stay in the
region $\wh\rho>2\wh m$.

Let us now put this solution in the background of the solution given in
\refb{e2.1} at the position $\rho$, taking the size of \refb{e2.1} to be much larger than
the size of this solution (i.e. we take $m,\rho >> \wh m, \wh \rho>>1$). Then we can use
\refb{e2.3} to write
\be
e^{-2\wh\phi_0} = m\, \rho^{-1}\, \cosh\beta\, .
\ee
Taking $\wh m \wh\rho^{-1} \sinh^2 \alpha$ to be large, 
we get, from \refb{efro1},
\refb{e2.1nnn}:
\be\label{efin12}
e^{2\phi_{10}} = {1\over 2}\, m^{-1}\, \rho \, (\cosh\beta)^{-1} \, \wh m^{-1}\, \wh\rho \, 
(\sinh\alpha)^{-2}\, , \qquad R_9 = {1\over \sqrt 2}\, \wh m^{-1/2}\, \wh\rho^{1/2} \, 
(\sinh\alpha)^{-1}\, .
\ee

We now put the first black hole, described in section \ref{sone}, at the position
$\wh\rho$ by taking $m,\rho >> \hat m, \hat\rho>>r_\pm, a,b>>1$. This amounts to
replacing $e^{2\phi_0}$ in \refb{e1.9} by $e^{2\phi_{10}}$ given in
\refb{efin12}. Also, the radius of the ninth circle will now be multiplied by $R_9$ given
in \refb{efin12} since this becomes the asymptotic value of this radius from the
perspective of the solution \refb{e1.9}. This gives the following solution:
\ben\label{e1.9nnn}
&& \wt R = 
\left( { r_++b\over b+a}\right)^{1/4}, \qquad
\wt R_9 = \left( { r_++b\over b+a}\right)^{1/4} 
{1\over \sqrt 2}\, \wh m^{-1/2}\, \wh\rho^{1/2} \, 
(\sinh\alpha)^{-1},
\nonumber \\ && e^{2\wt \phi_{10}} 
= {1\over 2}\, m^{-1}\, \rho\, (\cosh\beta)^{-1} \, \wh m^{-1}\, \wh\rho \, 
(\sinh\alpha)^{-2} \, \left( { r_++b\over b+a}\right)^{3/2}\, .
\een
Note that if we ignore the ${1\over 2}\, \wh m^{-1}\, \wh\rho \, 
(\sinh\alpha)^{-2}$ factor, then the solution reduces to \refb{e1.9a}.

The final step is to make a $R\to R^{-1}$ duality transformation on the ninth
circle to map this to a type IIB string theory. Denoting the new variables by bars,
we get
\ben
&& \overline{ R} = \wt R = \left( { r_++b\over b+a}\right)^{1/4}, \qquad
\overline{ R}_9 = \wt R_9^{-1} =
\left( { r_++b\over b+a}\right)^{-1/4} 
{\sqrt 2}\, \wh m^{1/2}\, \wh\rho^{-1/2} \, 
\sinh\alpha\, ,
\nonumber \\ && e^{2\overline{ \phi}_{10}} = e^{2\wt\phi_{10}} \wt R_9^{-2}
=  m^{-1}\, \rho\, (\cosh\beta)^{-1} \, \left( { r_++b\over b+a}\right)\,  .
\een
In order to get type IIB string theory with large compactification radii and weak or
moderate coupling, we need $\overline{ R}, \overline{ R}_9>>1$ and 
$e^{\overline{ \phi}_{10}}\lsim 1$.
If we choose $\rho\sim m$ and $\wh\rho\sim\wh m$, then this requires:
\be
r_+>>a,b, \qquad \sinh\alpha >>  \left({r_++b\over b+a}\right)^{1/4}, \qquad 
\cosh\beta \gsim \left( { r_++b\over b+a}\right)\, .
\ee
The analog of \refb{erangeiia} now takes the form:
\be \label{erangeiib}
m >> \wh m >> r_+ >>> \sinh^4\alpha >>
r_+/(b+a) \sim \cosh\beta >> 1,  \qquad \rho \sim m, \qquad \wh\rho\sim\wh m\, .
\ee

Finally, we note that if we had used the type IIB description from the begining, then the
first black hole would carry D1-brane charge along the ninth
direction, the second black hole would carry equal amount of fundamental string winding
and momentum charges along one of the circles and the third black hole would carry
momentum along the ninth direction.

\sectiono{Achieving the goals with single black holes} \label{ssingle}

In the constructions described above we made use of multiple black holes to produce
arbitrarily large regions  where a local observer experiences ten dimensional type IIA or
type IIB string 
theories or eleven dimensional M-theory. It is natural to ask if this can be 
achieved using a single black hole. In this section we can describe how this can be
done.

For this analysis we shall make use of black holes carrying charges
representing momenta along the compact directions.
We shall first describe the construction in type IIA string theory on $T^6$.
This can again be obtained as a special case of the solution given in
appendix \ref{selectric} 
where we set $\alpha=-\beta$ and take the vectors $\vec p=\vec n$ to be
some generic six dimensional unit vector. Then 
the four dimensional metric $ds_4^2$, the four dimensional 
dilaton $\phi_{4}$ and the
internal six dimensional metric $ds_6^2$ 
describing the moduli fields have the following functional dependence on the
radial variable $\rho$::
\ben\label{eapparent}
ds_4^2 &=& -{\rho-2m\over \rho+2m\sinh^2\alpha} dt^2 + {\rho\over \rho-2m} d\rho^2
+ \rho^2 d\Omega_2^2
\, , \nonumber \\
e^{2\phi_{4}} &=& e^{2\phi_0} \, 
(1+ 2 m \rho^{-1} \sinh^2\alpha)^{-1/2}, \nonumber \\ 
ds_6^4 &\equiv& G_{ij} dy^i dy^j=
 (R^2-1) (\vec n .{d\vec y})^2 +  {d\vec y}^{\, 2}, \qquad
R= (1 + 2 m \rho^{-1}  \sinh^2\alpha)^{1/2}\, ,
\een
where $\phi_0$ is the asymptotic value of the dilaton $\phi_{4}$. The horizon of
the solution is at $\rho=2m$ and the momentum 
charge vector carried by the black hole is
proportional to $m\, \sinh^2\alpha\, \vec n$. 
As before, we shall take the large $m$ limit at fixed $\rho/m$ so as to smoothen the 
geometry and make it locally flat.
From \refb{eapparent} 
we can also compute the ten dimensional dilaton
\be\label{etendil}
e^{2\phi_{10}} = e^{2\phi_{4}} \, R = e^{2\phi_0}\, .
\ee
Thus the ten dimensional dilaton is constant. This shows that this is a purely
gravitational solution and hence can be easily lifted to any theory, {\it e.g.}
M-theory on $T^7$ or type IIB on $T^6$.

By taking $\rho\sim m$ and $\alpha$ large, we can take $R$ to be large. 
In this case, 
apparently
\refb{eapparent} indicates that only one of the directions $\vec n$
of the torus becomes large
and the orthogonal directions remain finite.
For example, if each $y^i$ is periodic with period one, and if $\vec n\propto
(1,1,\cdots 1)$, then the cycle that shifts $y^1$ by $2\pi$ and $y^2$ by $-2\pi$
will have finite size since it will not be affected by the term proportional to $R^2$
in the metric. However
by choosing $\vec n$ appropriately we can ensure that  
all closed cycles on the torus 
have large size. We shall illustrate this by an example below. Let us take
\be\label{edefn}
\vec n = (1,c,c^2,\cdots c^{d-1}) / \sqrt{1+c^2+\cdots + c^{2d-2}}
\ee
where $d$ is the dimension of the torus (6 for tyoe II theories and 7 for M-theory)
and $c$ is a small number which we take to be of order $R^{-1/d}$.\footnote{$c$ is
a rational number but this does not prevent us from taking it to be small.}
Then the cycle that shifts $y^i$ by $2\pi m_i$ for
integers $m_i$ will have size
\be \label{ecycle}
2\pi \left[m_1^2 + \cdots + m_d^2 + (R^2-1) (m_1+m_2 c +\cdots m_d c^{d-1})^2
/ (1 + c^2 +\cdots +c^{2d-2})
\right]^{1/2} \, .
\ee
For generic $m_i$ the cycle will have size of order $R$. We could avoid it
by choosing the $m_i$'s so that $(m_1+m_2 c +\cdots m_d c^{d-1})$ is small.
This can be done
in two ways: either set some of the $m_i$'s to zero or choose the $m_i$'s such
that the different terms in the sum cancel. In the first approach the smallest
value is achieved by setting $m_i=0$ for $i\ne d-1$ and $m_d=1$. 
This gives a cycle size of
order $2\pi R  c^{(d-1)}\sim R^{1/d}$ for $c\sim R^{-1/d}$. Thus we have
a large size for large $R$. On the other hand in order to have cancellation between
different terms in $(m_1+m_2 c +\cdots m_d c^{d-1})$, some of the $m_i$'s must be
of order $c^{-1}$. Then the $R$ independent terms in \refb{ecycle} will have
large contribution and we still will have a minimum cycle size of order 
$R^{1/d}$.

One can also see this by analyzing the spectrum of momentum carrying states. If
$\vec k$ denotes the quantized momentum along $T^d$ taking value on a
lattice of integers, then for the metric given in \refb{eapparent}, its
contribution to the mass$^2$ in four space-time dimensions is given by:
\be\label{emass2}
m^2=G^{ij} k_i k_j =  \vec k^2 + (R^{-2}-1) (\vec n.\vec k)^2\, .
\ee
Let us for definiteness take
\be
R = N^d, \qquad c= N^{-1}\, ,
\ee
for some large integer $N$. This can be achieved by taking the momenta along the $d$
circles to be some multiple of $(1,N,N^2,\cdots N^{d-1})$.
We now introduce the following
basis states in the momentum lattice:
\be
\vec e_1=(1,0,\cdots 0), \quad \vec e_2=(N, 1, 0,\cdots 0), \quad 
\vec e_3=(N^2,N,1,0,\cdots 0),
\quad \cdots, \ \vec e_d = (N^{d-1}, N^{d-2}, \cdots 1)\, .
\ee
We also have from \refb{edefn}
\be
\vec n = (1+N^2+\cdots N^{2d-2})^{-1/2} (N^{d-1}, N^{d-2},\cdots 1)\,  .
\ee
Then a general momentum vector may be expressed as 
\be
\vec k = \sum_{i=1}^d q_i \vec e_i, \qquad q_i\in \ZZZ\, .
\ee
Substituting this into \refb{emass2}, we get
\be
m^2 = N^{-2} \sum_{i=1}^d q_i^2 +\OO(N^{-3})\, .
\ee
This agrees with the spectrum of a theory where each of the $d$ circles is compactified on a
circle of radius $N=c^{-1}$.

This achieves the goal of having large 
dimensions of the torus. This can be applied to each of the three theories.
In particular, by
regarding the original four dimensional theory as type IIA on $T^6$ and considering 
a black hole that carries large 
six dimensional momentum charge we can get type IIA in flat ten dimensional
space-time. The same theory can be regarded as M-theory on $T^7$ and now by
switching on appropriate large seven dimensional momentum charge (which 
corresponds to six dimensional momentum charge and D0-brane charge in the
original type IIA description) we can get
M-theory in flat eleven dimensional space-time. The same theory can 
also be regarded
as type IIB on $T^6$ via a T-duality transformation and by switching on 
large six dimensional momentum charge (which corresponds to momentum along
five of the circles and fundamental string 
winding along the sixth circle in the type IIA description) we
can get type IIB theory in flat ten dimensional space-time.

\sectiono{Discussions} \label{sthree}

In \cite{2502.07883} we had described a general strategy for producing almost locally flat
background where the moduli fields take a different value compared to the 
values they take at infinity. In this paper we have described a particular example of this
mechanism where starting from a four dimensional string theory with finite values of all
the moduli fields, we have described backgrounds that have large locally flat regions 
where the space-time is effectively 
ten dimensional and the string coupling remains finite and another background that
has a large 
locally flat region where
the space-time is effectively eleven dimensional.

This of course 
does not demonstrate  that we can reach all possible values of the moduli this way.
It will be interesting to try to establish this or find the limitations if any. 
In particular the moduli space of string
theory has interesting topology changing transitions and one would like to realize
them as part of space-time background. One interesting case of these
is the conifold transition\cite{9504090,9504145} -- 
that takes a deformed conifold to a resolved conifold. 
One could ask if it is possible to construct a background
where in some region of space-time the internal manifold is a resolved conifold and in another
region it is a deformed conifold. Since this involves both vector multiplet
and hypermultiplet moduli of $N=2$ supersymmetric string theory, the configuration is
likely to involve both black holes and loops of black strings.

Finally we note that while the asymptotic geometry is taken to be flat space-time (times an
internal manifold) the geometries that we can construct in the interior are not limited to flat
space-time. For example in ten dimensional type IIB string theory we can take a stack of
$N$ spherical 
D3-branes  so that it has finite energy, but take the
radius of the sphere to be very large so that the relaxation time is 
large.
In that case the
near horizon geometry of a local patch of the D3-branes will be 
$AdS_5\times S^5$ with five form flux\cite{9711200} and
we can study properties of this background by sending observers close to the
horizon.\footnote{If we scale the 4-form field strength according to \refb{escaling} then the
three-brane charge will grow as $\lambda^4$\cite{2502.07883} 
and we shall get a locally flat space-time.
To get $AdS_5\times S^5$ geometry we need to keep the three brane charge fixed.}
Of course if we try to approach infinitely close to the horizon then 
any experiment will take
infinite amount of time from the asymptotic observer's point of view due to the red shift, 
but once we fix the accuracy with which we want to do the 
experiment and estimate the time that will be needed from the asymptotic observer's
point of view to perform such an experiment, 
we can take the size of the 3-brane system to be sufficiently large so that
it does not evolve appreciably during the period of the experiment.
Similarly by taking large stacks of spherical M2-branes or M5-branes in M-theory
we can produce $AdS_4\times S^7$ or $AdS_7\times S^4$ geometries\cite{9711200}.
Since both, the 
ten dimensional type  IIB string theory and the eleven dimensional
M-theory, can be produced
as backgrounds in the interior of an asymptotically flat four dimensional string theory,
obtained by compactifying
type IIA string theory on $T^6$, we see that all the $AdS_p\times S^q$ backgrounds
listed above can be considered as different states of the same underlying theory.
In fact, in this case,  we could also wrap the branes along the compact cycles instead
of taking them to be spherical since from the asymptotic observer's viewpoint these
are particle like states carrying large but finite energy and charges.

\bigskip

\noindent{\bf Acknowledgement:} 
This work was supported by the ICTS-Infosys Madhava 
Chair Professorship
and the Department of Atomic Energy, Government of India, under project no. RTI4001.

\appendix

\sectiono{Electrically charged black hole solution in type II on $T^6$} \label{selectric}

In this appendix we shall use the results of \cite{9411187} to
describe the non-rotating
black hole solutions in type II string theory on $T^6$ carrying fundamental string
winding and momentum charges.\footnote{Even though the solution 
in \cite{9411187}
was given for heterotic string theory, the result in type II string theory can be obtained
easily from there by truncating the 28 dimensional charge vector to 12 dimensional 
charge vector and the $28\times 28$
matrix valued scalar moduli $M$ to a $12\times 12$ symmetric $SO(6,6)$
matrix. The solution given here is obtained from the one in \cite{9411187} by setting
to zero the rotation parameter $a$. Also $\Phi$ in 
\cite{9411187} is our $2\phi_4$.}
The four dimensional string metric $ds_4^2$,
the four dimensional  dilaton $\phi_{4}$  the gauge
fields $A_\mu$, and the moduli of $T^6$, described by a 
$12\times 12$ symmetric
$O(6,6)$ matrix $M$, take the form:
\ben \label{e2.1app}
ds_4^2 &=& -\Delta^{-1} \rho^2 (\rho^2-2m\rho) dt^2 + \rho^2 (\rho^2-2m\rho)^{-1} d\rho^2
+ \rho^2 d\Omega_2^2 \, , \nonumber \\
e^{-2\phi_{4}} &=& e^{-2\phi_0} \, \Delta^{1/2} \, \rho^{-2} \, , \nonumber \\
M &=& I_{12} + \pmatrix{Pnn^T & Q n p^T \cr Q p n^T & Ppp^T} \, , \nonumber \\
A_t &=& -{1\over \sqrt 2} \, m\, \rho\, \Delta^{-1} \, \pmatrix{
\sinh\alpha \{\cosh\beta \rho^2
+m\rho(\cosh\alpha-\cosh\beta)\} \vec n \cr
\sinh\beta \{\cosh\alpha \rho^2
+m\rho(\cosh\beta-\cosh\alpha)\} \vec p
}\, ,
\een
where
\ben \label{edeldef}
\Delta &\equiv& \rho^4 + 2 m\rho^3(\cosh\alpha\cosh\beta-1) + m^2 \rho^2 
(\cosh\alpha-\cosh\beta)^2  \, , \nonumber \\
P &\equiv & 2\, \Delta^{-1} \, m^2 \rho^2 \sinh^2\alpha\sinh^2\beta\, , \nonumber \\
Q &\equiv & - 2\Delta^{-1}m\rho  \sinh\alpha\sinh\beta \{ \rho^2 + m\rho (\cosh\alpha\cosh\beta
-1)\}
\, .
\een
$m$, $\alpha$, $\beta$ and the six dimensional unit vectors $\vec n$ and $\vec p$
are parameters labelling the mass and the twelve electric charges carried by the
black hole.
$\phi_0$ is the asymptotic value of the dilaton field that was not included in
\cite{9411187} but has been included here for convenience. $\vec n$ and $\vec p$
describe the directions of the
winding$\mp$momentum
charges along the six internal directions. $M$ encodes information
about the components of the metric and NSNS 2-form field along $T^6$. 
For our purpose it will be sufficient to note that if we have a diagonal metric
on $T^6$ where five of the circles have 
radius 1, the sixth circle has radius $R$ and the NSNS 2-form field vanishes,
then $M$ takes a block diagonal form, with identity matrix in the first 
$5\times 5$ block and the $5\times 5$ block spanning 7-11th rows and columns, 
and takes the following form in the 6-th and 12th rows and columns:
\be\label{emr}
{1\over 2} \pmatrix{R^2+R^{-2} & R^2-R^{-2} \cr R^2-R^{-2} & R^2+R^{-2}}\, .
\ee

\end{document}